\documentclass[12pt]{aastex}

\usepackage{emulateapj5}

\slugcomment{Accepted for publication in the FUSE Special Issue of ApJ Letters, May 10, 2000}

\newcommand\LL{\mbox{$\:\lambda\lambda $}}
\newcommand{\kms}{km~s$^{-1}$}
\newcommand\etal{et~al.}
\newcommand\Hone{H\,{\sc i}}
\newcommand\Arone{Ar\,{\sc i}}
  
\newcommand\Ctwo{C\,{\sc ii}}
\newcommand\Cthree{C\,{\sc iii}}
\newcommand\Cfour{C\,{\sc iv}}

\newcommand\Sfive{S\,{\sc v}}

\newcommand\None{N\,{\sc i}}

\newcommand\Oone{O\,{\sc i}}

\newcommand\Othree{O\,{\sc iii}}
\newcommand\Ofour{O\,{\sc iv}}
\newcommand\Ofive{O\,{\sc v}}
\newcommand\Osix{O\,{\sc vi}}

\newcommand\Mgtwo{Mg\,{\sc ii}}
\newcommand\Sitwo{Si\,{\sc ii}}
\newcommand\Sithree{Si\,{\sc iii}}
\newcommand\Sifour{Si\,{\sc iv}}
\newcommand\Fetwo{Fe\,{\sc ii}}
\newcommand\Ptwo{P\,{\sc ii}}  
\newcommand\Neeight{Ne\,{\sc viii}}
\newcommand\lam{$\lambda$}

\begin{document}

\title{FUSE Observations of the Galactic and 
Intergalactic Medium Towards H1821+643}

\author{W. R. Oegerle\altaffilmark{1}, 
T. M. Tripp\altaffilmark{2},
K. R. Sembach\altaffilmark{1},
E. B. Jenkins\altaffilmark{2},
D. V. Bowen\altaffilmark{2}, 
L. L. Cowie\altaffilmark{3},
R. F. Green\altaffilmark{4},
J. W. Kruk\altaffilmark{1},
B. D. Savage\altaffilmark{5}, 
J. M. Shull\altaffilmark{6},
D. G. York\altaffilmark{7}
}

\altaffiltext{1}{Department of Physics and Astronomy, Johns Hopkins University,
Baltimore, MD 21218}
\altaffiltext{2}{Department of Astrophysical Sciences, Princeton University,
Princeton, NJ 08544-1001}
\altaffiltext{3}{Institute for Astronomy, University of Hawaii, Honolulu, 
HI 96822}
\altaffiltext{4}{Kitt Peak National Observatory, NOAO, P.O. Box 26732, 
950 N. Cherry Ave, Tucson, AZ 85726-6732}
\altaffiltext{5}{Department of Astronomy, Washburn Observatory, University
of Wisconsin, Madison, WI 53706}
\altaffiltext{6}{CASA and JILA, Dept. of Astrophysical and Planetary Sciences, 
University of Colorado, Boulder, CO 80309}
\altaffiltext{7}{Department of Astronomy, University of Chicago, Chicago, IL}

\begin{abstract}

We have obtained moderate  resolution spectra of the bright QSO
H1821+643 ($z=0.297$) in the wavelength interval $990-1185$\AA\/ with
FUSE.   Strong, complex \Osix\ \lam1032 and \Fetwo\ \lam1145 absorption
arising in gas above several spiral arms and the outer warp of our
Galaxy is detected.  We have identified absorption by a high
negative-velocity ($-215$ \kms\/) component of \Osix\/, which
corresponds to the limiting velocity of  $v_{LSR}\sim-190\pm20$ \kms\/  
for Milky Way gas along this line of sight, assuming corotation of disk
and halo gas and a flat rotation curve for the outer galaxy.  We report
the detection of four absorbers in the intergalactic medium, through
detections of Ly$\beta$ at $z=0.2454$ and $z=0.1213$, Ly$\delta,
\epsilon, \zeta$ in a system at $z=0.2249$, \Cthree\ \lam977\ at
$z=0.1705$, and \Osix\ \lam1032\ at $z=0.12137$.  The FUSE data show
that the Lyman absorbers at $z=0.225$ are composed of 2 components
separated by $\sim 70$ \kms\/. Finally, we have observed associated
absorption from \Othree\ \lam832, \Ofour\ \lam787, and possibly \Sfive\
\lam786 at the redshift of the QSO ($z\sim 0.297$). \Neeight \LL770,780
is not detected in this associated system.   When combined with the
previous detection of  associated \Osix\ \lam1032 \citep{savage98}, we
conclude that this system is a multiphase absorber, and we discuss the
origin of the absorption in this context.

\end{abstract}


\section{Introduction}

In this paper, we present intermediate resolution ($\sim 20$ \kms\/)
spectra of the quasar H1821+643 ($V=14.2, z=0.297$) which is visually
the brightest known object in the sky at $z > 0.2$.  H1821+643 is an X-ray
selected, radio-quiet quasar  \citep{pravdo84}, and resides in a cD
galaxy at the center of a rich cluster of galaxies \citep{schneider92}. 
The optical properties of the QSO have been reported by \citet{hutch91}
and \citet{kolman93}.  The latter authors  find evidence for a strong
optical/UV bump and a soft x-ray excess. The cluster of galaxies hosting
the QSO is one of the most luminous X-ray clusters known \citep{hall97}.
Spectroscopic studies of H1821+643 have been carried out in the UV at
low resolution \citep{bahcall92,kolman93,tripp98} and intermediate
resolution \citep{savage95,savage98,penton00}, in order to study  the
Galactic and intergalactic medium along the line of sight.

\section{Observations}

H1821+643 was observed by FUSE on Oct 10 and 13, 1999 for a total
integration time of 48.8 ksec. \citet{moos00} and \citet{sahnow00}
provide an overview of the FUSE instrument and its performance.   Data
in the wavelength interval $990-1185$  \AA\/ covered by both LiF
channels were obtained.  Processing steps included orbital doppler
compensation, background subtraction, flux and wavelength calibration. 
No flat-fielding of the data was performed. The dispersion solution was
derived from pre-launch calibration spectra, and adjusted based on
comparison of FUSE absorption line velocities to velocities of
appropriate absorption lines in the STIS/HST echelle spectrum of
H1821+643 obtained by \citet{tripp00}. In general, the velocities are
accurate to $\sim 5$ \kms\/.  The measured spectral resolution was $\sim
20-25$ \kms\ depending on wavelength.   The detector background and
scattered light in the  instrument are extremely low, leading to
accurate knowledge of the residual intensities in absorption lines.
Equivalent widths  and column densities of selected absorption lines
were measured using  the techniques described by \citet{sembach92} and
are reported in Table 1.  Spectra shown in Figures 1-3 have been
smoothed by 5 pixels (0.03\AA\/).

\section{Galactic Absorption}

The H1821+643 sight line $(l,b) = (94^\circ, 27^\circ)$  passes over 3
spiral arms seen in \Hone\ emission in a survey above the galactic
plane by \citet{kepner70}: (1) the intermediate arm
($v_{LSR}\sim -42$ \kms\/), the  Perseus arm ($v_{LSR} \sim -75$ \kms\/)
and the outer arm ($v_{LSR} \sim -93, -120$ \kms\/). For the
observations reported here $v_{LSR} = v_{hel} + 16$ \kms\/.   The sight
line is also in a direction where the warp of the outer galaxy extends
to large Galactic latitudes. The interstellar spectrum of H1821+643 is
dominated by lines of H$_2$ and \Fetwo\/ in the FUSE bandpass.   Also
present are \Arone\/, \Ctwo\/, \Sitwo\/,  \Ptwo\/, and \Osix\/. Strong
\Oone\ and \None\ lines were also detected in spectra obtained while the
satellite was in the  earth's shadow, where the terrestrial day glow
emission lines in these species are absent. The $1030-1040$\AA\ region
of the spectrum displaying \Osix\ \LL1032,1038, \Ctwo\ \lam1036 and
several H$_2$ lines is shown in Figure 1.  

\begin{figure*}
\plotone{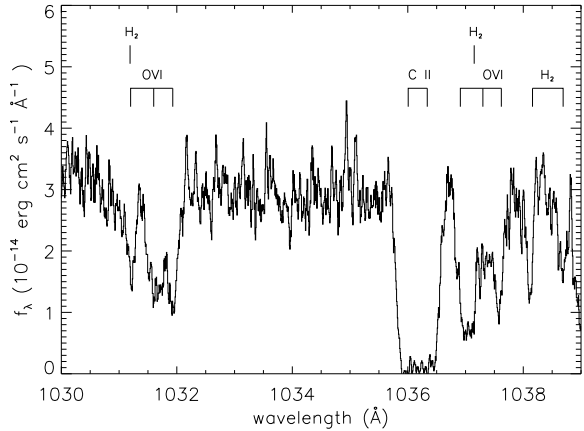}
\caption{The 1030-1039\AA\ region of the FUSE spectrum showing the
complex \Osix\/, broad \Ctwo\ \lam1036\/, and
numerous H$_2$ Galactic absorption lines.}
\end{figure*}

The \Fetwo\ lines are strong and have negative velocity wings
extending out to as much as $\sim -170$ \kms\/ for \Fetwo\ \lam1145. 
These negative velocity components are formed above the outer arm, and
are also seen in \Mgtwo\ and \Sitwo\ lines in GHRS data \citep{savage95}.
The H$_2$ lines are local and are not seen at large negative velocities.

It is interesting to compare the strengths  of the \Arone\ \LL1048,1066
lines to those in the \None\ multiplet at 1134\AA\/.  In a fully neutral
medium, these lines should have similar equivalent widths since they
have nearly similar values of $P \equiv {\rm log}(A) + {\rm
log}(f\lambda)$ where $A =$ cosmic abundance, i.e., $P \sim 8.5-9.0$.
However, the \Arone\ lines are noticeably weaker than the strongly
saturated \None\ lines in the spectra of H1821+643.  \citet{sofia98}
have pointed out that in diffuse clouds \Arone\ is unlikely to be
depleted onto dust grains.  Nevertheless, they argue that in regions
that are partially ionized by EUV radiation, \Arone\ may appear to be
deficient relative to \Hone\ (or \None\/) because it has a substantially
larger ionization cross section and thus might be more ionized.  Note
that \Arone\ is only clearly detected in the intermediate arm. It is not
detected in the Perseus or outer arms, probably because the Ar is more
highly ionized at greater distances from the Galactic plane.

The \Osix\ absorption at $-120$ to $-150$ \kms\ is associated with the
outer arm and distant warp of our Galaxy.  If we assume that the halo 
corotates with the underlying  disk, the implied Galactocentric distance
of the  absorbing gas is $\sim 25-50$ kpc, and the distance above the
plane is $\sim 10-20$ kpc.  Independent evidence for the existence of
\Osix\ absorption in the outer halo of our galaxy is supplied by the
{\it absence} of \Osix\ absorption at velocities exceeding $-70$ \kms\
in the FUSE spectrum of K1-16, which is only $85''$ away from H1821+643
on the sky, and is at a distance of 1.6 kpc \citep{kruk00}. 
Consequently, the high negative-velocity \Osix\ absorption seen in
H1821+643 must be formed in the Galactic halo beyond K1-16.

A high negative-velocity ($-215$ \kms\/) component of \Osix\ \lam1031.93
is coincident with a low velocity H$_2$ line from the $(6-0) P(3)$
rotational level at 1031.19\AA\/.  We have modeled the $J=3$ H$_2$ lines
in the spectrum, and find a good fit for a total column density of
N(H$_2$) $\sim 2 \times 10^{15}$ cm$^{-2}$ and a cloud excitation
temperature of $T_{ex} \sim 500$ K.  Based on this model, the expected
$H_2$ absorption at 1031.19\AA\ cannot account for the depth or width of
the observed line, indicating significant absorption by \Osix\ \lam1032
at $-215$ \kms\/. This confirms the tentative identification of this
component in the \Cfour\ \lam1549 line made with GHRS/HST by
\citet{savage95}.  This detection is quite interesting, since the line
velocity corresponds to the limiting velocity of $v_{LSR}\sim-190\pm20$
\kms\ for Milky Way gas assuming corotation and a flat rotation curve for the
outer galaxy.  Collisional ionization seems the most likely source of
\Osix\ ionization in this component.  Photoionization would require a
very hard radiation field ($E>114$ eV) to ionize \Ofive\ to \Osix\/, and
the large photoionization parameter, $U$, requires a very low value of
$n_H$ and an extremely long pathlength, $l=N_H/n_H>100$ kpc, to produce the
absorption (see \citet{sembach00}).

\section{Intervening Absorption}

\citet{tripp98} have detected strong (W$_\lambda > 200$ m\AA\/)
Ly$\alpha$ towards H1821+643 at redshifts of 0.1213, 0.1476, 0.1699,
0.2132 and 0.2249. \citet{savage95} also reported intervening Ly$\alpha$
absorption at $z=0.02454$.  We have detected absorption from four of 
these absorbers in the FUSE spectrum, and the column densities are
reported in Table 1.  The  Lyman series lines detected with FUSE in
these systems provide  important constraints on N(\Hone\/) since the
stronger lines  in the HST bandpass are saturated.  The reader is also
referred to \citet{shull00} for discussion of intervening Ly$\beta$
absorbers.  One must be aware of the potential for
ejected material from the QSO to masquerade as intervening gas, as 
shown for \Cfour\ absorption line systems by \citet{richards99}.
We believe that there is good evidence that the absorbers in H1821+643
are truly intervening, based on the proximity of intervening galaxies to 
the line of sight \citep{tripp98,bowen00}.

\citet{tripp98} report a strong, blended line of Ly$\alpha$ at
$z=0.12123, 0.12157$. In addition to detecting the the Ly$\beta$ line in
this system, we report the detection of a line at 1157.17\AA\ which we
have identified as \Osix\ \lam1032\/ at $z=0.12137$ (see Figure 2).  The
line is clearly seen in both the LiF1 and LiF2 spectra. The \Osix\
\lam1038\ line is blended with the Ly$\delta$ line in the $z=0.225$
absorber and could not be measured.  This is the sixth {\it intervening}
\Osix\ absorption system observed towards H1821+643, and provides
further evidence that low-$z$ \Osix\ systems contain a large fraction of
the baryons at the present epoch \citep{tripp00,tripp00_2}.  These
observations are in accord with cosmological simulations by
\citet{cen99} that predict a substantial fraction of present-day baryons
are in a shock-heated phase at  $10^5-10^7$K.

We have identified an observed line at 1143.6\AA\ as \Cthree\ \lam977 at
$z=0.1705$. This identification was made possible by comparison with the
wavelength of a Ly$\alpha$ absorber recently observed by \citet{tripp00}
with STIS/HST.  The Ly$\gamma$ line for this absorber is not
conclusively detected in the FUSE spectrum, although a weak, rather
broad feature is at the predicted wavelength.  The Ly$\beta$ line in
this absorber is redshifted into the strong Galactic \None\ triplet at
1200\AA\/. 

Intervening absorption at $z=0.225$ has been detected in the Ly$\alpha$
and Ly$\beta$ lines and in \Osix\ with HST \citep{savage98}.  Recently,
\citet{tripp00} have also clearly detected \Sithree\ in this system, and
the component structure establishes that this is a multiphase absorber.
However, this absorber is not detected by HST in low ionization lines
such as \Sitwo\ or the high ionization lines of \Sifour\ and \Cfour\/. 
\citet{savage98} showed that the absorption could occur in low density,
extended gas photoionized by the UV background or in hot collisionally
ionized gas in an intervening galaxy or galaxy group. Evidence for the
presence of a galaxy group at $z=0.225$ has been provided by
\citet{schneider92} and \citet{tripp98}.

The Ly$\delta$, Ly$\epsilon$ and Ly$\zeta$ lines at $z=0.22491$ are
detected in the FUSE spectrum.  Two of these lines are shown in Figure 2. The
lines are clearly resolved into 2 components with velocity separation of
$\sim 70$ \kms\/.  We have not identified any metal lines arising in the
$z=0.225$ system in the FUSE spectrum.  It would be interesting to
obtain FUSE short-wavelength (SiC) spectra covering $900-995$\AA\/,
where we have the possibility of detecting the redshifted \Neeight\
\LL770,780 doublet.  \Neeight\ has a higher ionization potential than
\Osix\ (207 eV vs. 114 eV).  If detected, it would indicate the presence
of collisionally ionized gas at a temperature of $T\sim10^{5.5}$K.

\begin{figure*}
\plotone{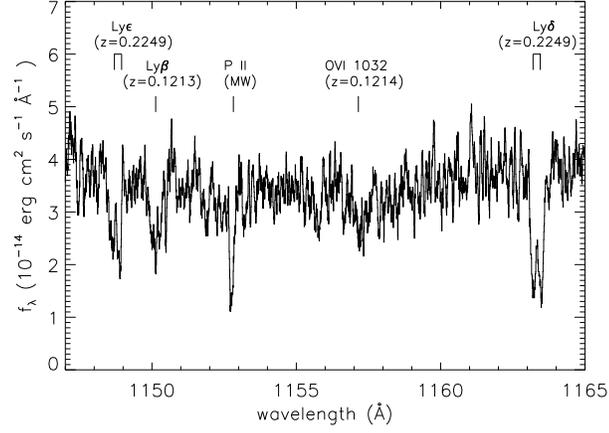}
\caption{The Ly$\delta$ and Ly$\epsilon$ intervening absorbers at
$z=0.225$, and the Ly$\beta$ and \Osix\ absorber at $z=0.1213$.  Milky Way (MW)
\Ptwo \lam1152 is also indicated.}
\end{figure*}

\section{Associated Absorption}

We report the detection of several ``associated'' ($z_{abs} \approx
z_{em}$) absorption lines  in the spectrum of H1821+643.  We identify
the observed line at $1021.45$\AA\/ as absorption by the rest-frame EUV
line \Ofour\ \lam787.7 (see Figure 3).  This line has a FWHM $\sim 150$
\kms\/,  which places it in the category of narrow absorption line (NAL)
absorbers \citep{weymann79,hamann00}. We have also identified \Othree\
\lam832.93 with an observed line at $\lambda_{obs} \sim 1080.05$\AA\/. 
Finally, we have tentatively identified a weak line at $\lambda_{obs}
\sim 1019.85$\AA\/ as \Sfive\  \lam786\/, although this line is much
narrower than the \Ofour\ \lam787 line.  These lines cannot be redshifted, 
higher order Lyman lines, because the corresponding Ly$\alpha$ lines
are not detected in the longer wavelength UV spectra obtained with GHRS/STIS.

Associated absorption by \Cthree\ \lam977, \Sifour\ \lam1394, \Cfour\
\LL1548,1550, \Osix\ \LL1032,1038 as well as several members of the Lyman
series of hydrogen have been detected in H1821+643 by \citet{savage98} and
\citet{penton00}. 
The FUSE and HST observations combined show a broad range of ionization
in the associated absorber(s) -- including \Othree\/, \Ofour\/, and
\Osix\/.  However, no low ionization species such as \Mgtwo\ or \Sitwo\
has been observed. We see no evidence for \Neeight\ \LL770,780
absorption in the FUSE data at $z_{abs}=0.29673$. 

\begin{figure*}
\plotone{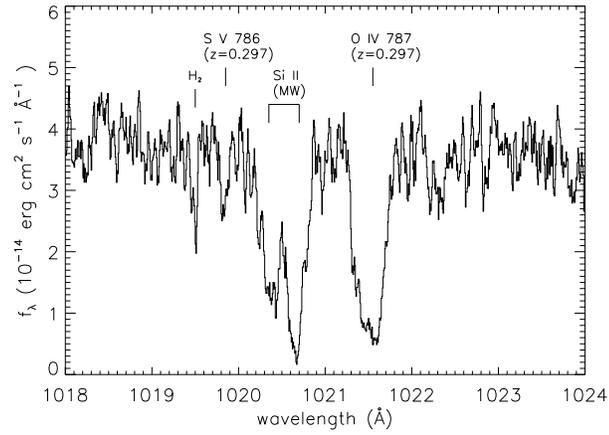}
\caption{\Ofour\ \lam787 and \Sfive \lam786 in the associated absorber
at $z=0.29673$ are indicated, as well as a multi-component Milky Way
\Sitwo\ \lam1020 line.}
\end{figure*}

There are 3 possible sites of associated absorption in the H1821+643
spectrum: (1) the host galaxy of the QSO, (2) the intracluster medium
(ICM), or (3) a cluster galaxy or galaxies along the line of sight to
the QSO.  Absorbing gas which is near the QSO central engine is often
called ``intrinsic'' absorption.  In many specific cases, there is
strong spectroscopic evidence that associated NALs are intrinsic.  The
evidence includes: (1) time variable line strengths, (2) smooth
absorption that is broad compared to the thermal line width, (3) partial
covering of the continuum source, and (4) presence of excited-state
absorption \citep{hamann00}.  None of these properties convincingly
describes the H1821+643 associated absorber.  Time-variability of
absorption lines has not been reported, and cannot be addressed with
this dataset.  The \Ofour\ line is broader than its thermal line width,
but is narrower than most intrinsic systems, which typically have widths
of $\sim 500$ \kms\/.   The associated Ly$\alpha$ line observed in the
HST/STIS spectrum obtained by \citet{tripp00} is black at the line
center (T. M. Tripp,  private communication). This requires full
coverage of the continuum  source. Finally, excited-state absorption is
not observed.  All of this evidence points to the associated absorber in
H1821+643 being unlike normal ``intrinsic'' absorbers.

Nevertheless, we still think that a likely origin of the associated 
absorption lies in the nearby environs of the H1821+643 host galaxy.
Halos of cluster galaxies will be ram-pressure stripped by their passage
through the intracluster medium.  Gas in the ICM is extremely hot 
($T\sim10^7$K), and will have negligible ionic fractions of \Othree\ and
\Ofour\/. If a cooling flow exists in the cluster, then the \Osix\ and
less ionized species formed in this cooling gas would exist close to the
cD galaxy at the cluster center -- the host of H1821+643.

The photoionization models presented by \citet{hamann97} indicate that
an ionization parameter of $U \sim 0.1$ (for a wide range of continuum
shapes) is needed to simultaneously produce \Ofour\ and \Osix\/, but
then the fractional ionization of \Othree\ will be much too low to
explain the absorption we detect.  Hence, a multiphase model is required
to explain the broad range of ionization present in the H1821+643
absorber. Multiple sites for the formation of the absorption is also
consistent with the \Ofour\ \lam787 line width.  

\acknowledgments

This work is based on data obtained for the Guaranteed Time
Team by the NASA-CNES-CSA FUSE mission operated by the
Johns Hopkins University. Financial support to U. S.
participants has been provided by NASA contract NAS5-32985.

\begin{deluxetable}{cllccl}
\tabletypesize{\scriptsize}
\tablecolumns{6}
\tablewidth{0pt}
\tablecaption{Equivalent Widths and Integrated Column 
Densities of Selected Absorption Lines}
\tablehead{
\colhead{$\lambda$\tablenotemark{a}} &
\colhead{Identification} &
\colhead{$z_{\rm abs}$}  &
\colhead{$W_{\rm r}$\tablenotemark{b}} & 
\colhead{$N_{\rm a}$\tablenotemark{c}} & 
\colhead{($v_{-}$,$v_{+}$)} \\
\colhead{(\AA )} & \colhead{} & \colhead{} & \colhead{(m\AA )} &
\colhead{($10^{14}$ cm$^{-2}$)} & \colhead{(km s$^{-1}$)}
} 
\startdata
998.99  & \Neeight\ $\lambda$770.41 & 0.2967 & $< 89$     & $< 1.8$\tablenotemark{d,e} & ($-90,90$) \\
1020.48 & \Sitwo\ $\lambda$1020.70 & 0.0     & 387$\pm$13 & ...\tablenotemark{f,g}        & 
($-$190,30) \\
1021.45 & \Ofour\  $\lambda$787.71  & 0.29673 & 239$\pm$9 & 7.6$\pm$0.5\tablenotemark{e} & 
($-$90,90) \\
1031.71 & \Osix\  $\lambda$1031.92 & 0.0     & 328$\pm$18 & 4.1$\pm$0.4 & 
($-$190,90) \\
1048.11 & \Arone\ $\lambda$1048.22 & 0.0     & 148$\pm$9 & $> 1.3$\tablenotemark{h}  & 
($-$80,10) \\
1050.73 & \Hone\ $\lambda$1025.72  & 0.02438 & 75$\pm$7  & 1.4$\pm$0.2 &
($-30$,25) \\
1055.09 & \Fetwo\ $\lambda$1055.26 & 0.0     & 103$\pm$20 & $>24.4$\tablenotemark{g,h} & 
($-$190,30) \\
1066.55 & \Arone\ $\lambda$1066.66 & 0.0     & 107$\pm$10 & $> 3.0$\tablenotemark{h}  & 
($-$80,10) \\
1080.05 & \Othree\  $\lambda$832.93  & 0.29669 & 84$\pm$16  & 1.8$\pm$0.3\tablenotemark{e} & 
($-$90,90) \\
1125.26 & \Fetwo\   $\lambda$1125.45 & 0.0     & 229$\pm$18 & $>21.2$\tablenotemark{g,h}  & 
($-$190,30) \\ 
1140.04 & \Hone\  $\lambda$930.75  & 0.22486   & 99$\pm$16   & 37.4$\pm$5.6 &
($-95$,70) \\
1143.60 & \Cthree\ $\lambda$977.02 & 0.17050  & 42$\pm$8   & 0.1$\pm$0.02 &
($-30$,20) \\
1144.74 & \Fetwo\  $\lambda$1144.94 & 0.0     & 439$\pm$19 & $>7.2$\tablenotemark{g,h}   & 
($-$190,30) \\
1148.72 & \Hone\   $\lambda$937.80  & 0.22491 & 135$\pm$16 & 35.2$\pm$6.1 & 
($-$95,70) \\
1150.13 & \Hone\   $\lambda$1025.72 & 0.12129 & 137$\pm$23 &  3.2$\pm$0.6  &
($-90$,120) \\
1152.70 & \Ptwo\  $\lambda$1152.82 & 0.0     & 127$\pm$14 & 0.8$\pm$0.1 & 
($-$80,10) \\
1157.17 & \Osix\ $\lambda$1031.92 & 0.12137 & 98$\pm$21 & 1.0$\pm$0.2 &
($-$100,150) \\
1163.35 & \Hone\ $\lambda$949.74  & 0.22491 & 230$\pm$14 & 34.5$\pm$3.9 & 
($-$95,70) \\ 
\enddata
\tablenotetext{a}{Vacuum heliocentric wavelength of the line centroid after 
corrections of the wavelength scale described in the text.}
\tablenotetext{b}{Restframe equivalent width integrated from $v_{-}$ to 
$v_{+}$.}
\tablenotetext{c}{Apparent column density $N_{\rm a} = \int N_{\rm a}(v) dv$
integrated from $v_{-}$ to $v_{+}$. Oscillator strengths are from 
\citet{morton91} unless otherwise indicated.}
\tablenotetext{d}{$4\sigma$ upper limit. }
\tablenotetext{e}{Oscillator strength from \citet{verner94}.}
\tablenotetext{f}{A lower limit has not been placed on $N$(\Sitwo\/) due to
possible blending with \Othree\ $\lambda$832.93 at $z_{\rm abs}$ = 0.225.}
\tablenotetext{g}{Oscillator strength from Morton (2000, unpublished):
$f$(\Sitwo\ 1020.70\/)$=0.0164$, 
$f$(\Fetwo\ 1055.26\/)$=0.00615$, 
$f$(\Fetwo\ 1125.45\/)$=0.0156$, 
$f$(\Fetwo\ 1144.94\/)$=0.109$. }
\tablenotetext{h}{Saturated absorption line.}
\end{deluxetable}

\end{document}